\documentstyle [12pt,epsfig] {article}
\title{
Application of Bent Crystals At IHEP 70-GeV Accelerator To Enhance The
Efficiency Of Its Usage
       }
\author{
V.I.Kotov, A.G.Afonin, V.M.Biryukov, Y.A.Chesnokov, N.A.Galyaev,\\
V.N.Gres, V.A.Maisheev, V.A.Medvedev,A.V.Minchenko,V.I.Terekhov,\\
E.F.Troyanov,Y.S.Fedotov, V.N.Zapolsky,B.A.Zelenov\\
{\em Institute for High Energy Physics, Protvino, Russia}\\
Yu.M.Ivanov, {\em PNPI, St.Petersburg, Russia};
W.Scandale, {\em CERN, Geneva}
}
\date{EPAC 2000 Proc. (Vienna), p.364}
\begin{document}
\maketitle

\begin{abstract}
Bent crystal was extracting 70-GeV protons with average intensity
4$\times$10$^{11}$
(as measured in external beamline)
per spill of $\sim$1.6 s duration, in parallel to the
simultaneous work of two internal targets in the accelerator ring.
An additional crystal, placed in the external beamline, was deflecting
a small part of the extracted beam with intensity $\sim$10$^7$ protons toward
another physics experiment. Crystal-extracted beam had a typical size of
4 mm by 4 mm {\em fwhm} at the end of the external beamline. Measurements
for the extraction efficiency and other characteristics at the
simultaneous work of four experimental set-ups are presented.
With crystal working in the above-said
regime during one month, no degradation of channeling was observed.
The studies of extraction efficiency have been continued with new crystals.
\end{abstract}

\section{Studies of beam extraction by means of bent crystal}

Since 1997 at the IHEP 70 GeV accelerator we carried out studies of proton
beam extraction by means of bent silicon crystals [1-4].
These studies pursued the realization of a pure multiturn extraction
using short, 5 to 3 mm, O-shaped crystals (see Fig. 1) with small bendings
of 2 to 0.5 mrad.
Such a small bending of a crystal is insufficient for direct extraction
of the beam out of accelerator. Therefore, crystals served as primary
element in the existing scheme of slow extraction.

\begin{figure}[h]
\parbox[c]{7cm}{\epsfig{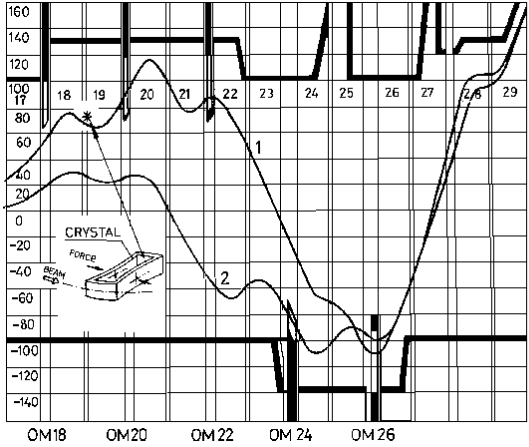}}

	\caption{
	The scheme of the accelerated-beam extraction via
	septum-magnets OM-20 (curve 1) and OM-24 (curve 2).
	The inset shows O-shaped crystal.
}
	\end{figure}

Crystals with bendings of 1.5 to 2 mrad were installed in straight section
19 of the accelerator before the septum-magnet OM-20 of the slow extraction
system and provided a kick of the deflected beam into the aperture of this
magnet with partition thickness of 8 mm (Fig. 1).

Another series of crystals with bendings of 0.5-1 mrad were installed
in straight section 106 (not shown in Fig. 1); here the extraction
of the crystal-deflected beam was performed through the septum-magnet OM-24
with partition thickness 2.5 mm (Fig. 1, curve 2).

\begin{figure}[h]
\parbox[c]{7cm}{\epsfig{file=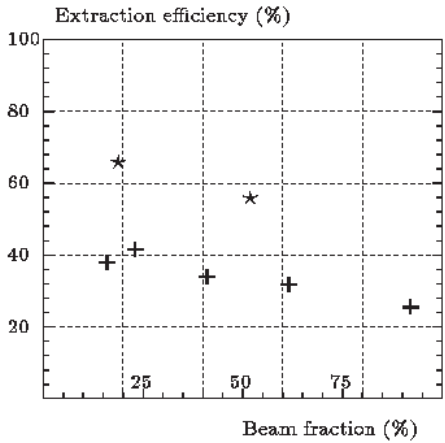,height=6cm}}
	\caption{
	Spill-averaged efficiency of extraction as measured with
	 3-mm crystal 0.9 mrad bent ({$\star$}), March 2000;
	 5-mm crystal 1.5 mrad bent ($+$), March 1998;
	plotted against the beam fraction taken from the accelerator.
}
     \end{figure}

To direct the accelerated beam onto the crystals,
the local orbit distortion systems were used
working in the regime of close loop operation.
The obtained efficiencies are plotted in Fig. 2.
The accuracy of the beam extraction efficiency measurements
was about 4\%.
For 20\% of the accelerator beam intensity dumped onto the crystal,
the measured extraction efficencies were 42\% for the 5-mm crystal
with 1.5 mrad bending and 65\% for the 3-mm crystal
with 0.9 mrad bending, installed in straight sections 19 and 106 respectively.
The reduction of extraction efficency with increase of the beam fraction
dumped onto the crystal is due to the drift of the beam angle at a crystal
because of the tilted phase ellipse.
During the studies of the extraction efficiency,
for a short period of time ($\sim$1 hour) the regimes
of beam extraction of intensity $\sim$1.2$\cdot$$10^{12}$
protons per cycle were exploited.

\section{Example of practical usage of bent crystals at IHEP accelerator}

The obtained high efficiencies of the beam extraction
with use of bent crystals open new ways for setting up  new
physical experiments at accelerators.
As an example, let us consider one of the schemes for the practical usage
of bent crystals realized at U-70 accelerator and shown in Fig. 3.

\begin{figure} [h]
\parbox[c]{7cm}{\epsfig{file=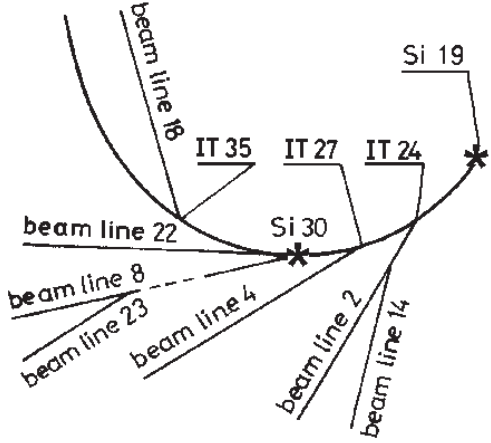,height=6cm}}

\caption{
One of the schemes of practical usage
of bent crystals realized at U-70 accelerator.
}
	\end{figure}

By means of 1.5 mrad bent crystal Si19 (straight section 19)
the proton beam of up to 4$\times$10$^{11}$ p/cycle intensity
was extracted into beamline 8 and directed into beamline 23 onto
the target of the experimental set-up used for the studies of
$K^{\pm}$-mesons decay modes into three $\pi$-mesons:
$K^{\pm}\rightarrow \pi^{\pm}\pi^0\pi^0$.
As the $K^+$-meson yield is about 20 times higher than that of $K^-$,
with the transfer from $K^-$ to $K^+$
the intensity of the extracted proton beam
had to be reduced by the same factor.
This procedure was repeated regularly with 24 hours period.

Along the path of the extracted beam, at the beginning of beamline 8
another crystal Si30 bent $\sim$9 mrad was positioned to deflect
a small fraction of the beam (up to 10$^7$ protons per cycle)
into beamline 22.
Simultaneously with crystals Si19 and Si30, two internal targets
IT24 and IT27 were working, supplying negatively-charged beams of
up to 10$^7$ particles/cycle intensity into beamlines 2 and 4 respectively.
Besides the physical studies on these four beamlines  (22, 23, 4, and 2),
methodical work was feasible on the test beamline 18.

To direct the accelerated beam onto crystal Si19
and two internal targets IT24 and IT27 along the flattop of magnet cycle
1.6 s long,
three local orbit distortion systems were used
in the regime of close loop operation.
\begin{figure}
\parbox[c]{7cm}{\epsfig{file=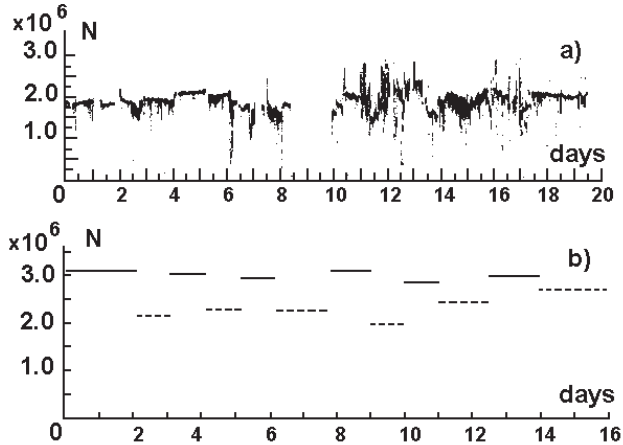,height=6cm}}

	\caption{
The intensity  N of secondary beams in beamline 2 (a)
and in the set-up at beamline 23 (b)
in the $K^-$ (solid) and $K^+$ (dashed) work regimes
}
	\end{figure}

Fig. 4 shows the intensity of secondary beams in beamline 2
and in the set-up at beamline 23 in the $K^+$ and $K^-$ work regimes.
For a time  about 27 days, the Si19 crystal has obtained an irradiation of
the order of 10$^{20}$ proton/cm$^2$.
Herein the crystal channeling properties have not changed,
as comes directly from the measurements of beam extraction efficencies
in the beginning (44$\pm$2 \%) and in the end (43$\pm$2 \%) of the run.

The first experience of the work with the above-considered system,
although satisfied the physicists, has also revealed  ways for
further advancement (improvement of the local orbit distortion systems
stability, possibility to increase the intensity 
in the $K^+$ and $K^-$ regimes on the set-up of beamline 23 etc.).

\section{Further plans}

As computations show, the efficiency of beam extraction can be raised
with use of shorter (order of 1 mm and less) bent crystals.

\begin{figure} [h]
\parbox[c]{7cm}{\epsfig{file=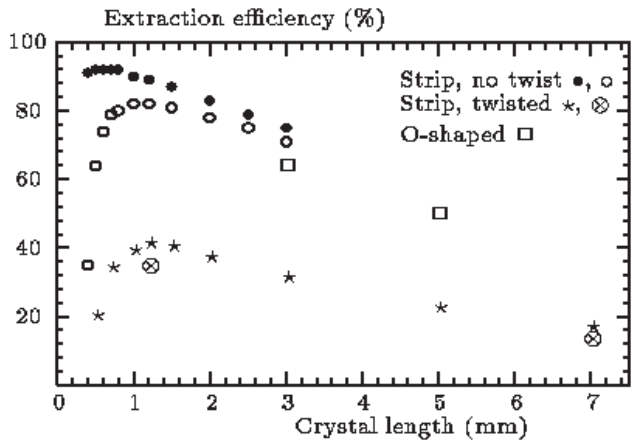,height=6cm}}
\caption{
Simulated efficiency of 70-GeV proton extraction by a crystal "strip"
40 mm high, 0.27 mm thick, with bending of 0.9 mrad.
No twist ($\bullet$, o); either perfect bending ($\bullet$) or
bending over half of the full length (o).
Below are shown ($\star$) the results with twisted crystal
(6 $\mu$rad/mm$^2$).
Also shown are the experimental results of 1997-2000 with
strips ($\otimes$), 1.2 mm and 7 mm long,
and O-shaped crystals ($\Box$), 3 mm and 5 mm long.
}

	\end{figure}

Besides the high energy accelerators,
crystals of this kind could be used also at low energy accelerators.
Basic problem here is to bend such a crystal at an angle of 0.5-1.5 mrad.
One of radical solutions for this problem is
to obtain bent crystals in the process of crystal growth \cite{5}.
Another direction, now in progress in IHEP, is related to the methodics
of bending a short strip.

The above said is illustrated by Fig. 5 where plotted are the calculations
of extraction efficiency for the most interesting, still-unstudied region,
and the experimental data obtained at the IHEP accelerator.

\end{document}